\documentclass{iopconfser}

\usepackage{graphicx}
\usepackage{float}
\usepackage{amsmath}

\begin{document}

\title{Trace dynamics, octonions and unification: An $E_8 \times E_8$ theory of unification}

\author{Tejinder P. Singh$^{1}$}

\affil{$^1$ Inter-University Center for Astronomy and Astrophysics, Pune, India}

\email{tejinder.singh@iucaa.in, tpsingh@tifr.res.in}

\begin{abstract}
This is a very brief overview of the ongoing research program of unification known as the octonionic theory. We highlight the quantum foundational origins for the theory, and the seven key ingredients which go into its making.
\end{abstract}

\section{A pre-spacetime, pre-quantum theory}
We have proposed a pre-spacetime, pre-quantum theory from which spacetime, gravitation, and quantum theory are emergent.
This theory is a (deterministic, non-unitary, and norm-preserving) matrix-valued Lagrangian dynamics.
From here, spontaneous localisation [precipitated by adequate entanglement] leads to the emergence of classical material bodies, spacetime, and gravitation.
On this background, the remaining pre-quantum degrees of freedom, upon coarse-graining:
    (i) obey quantum theory when the non-unitary aspect is negligible, or
    (ii) undergo quantum-to-classical transition, when non-unitarity is significant.  
This is also a theory of unification, from which standard model and general relativity emerge \cite{Singh1}.

\subsection{Motivation for this theory}
Quantum theory takes classical spacetime as a given.
However, classical spacetime is produced by classical material bodies.
These classical objects are a limiting case of quantum theory.
Therefore, quantum theory depends on its own limit: this is an approximation.
In a quantum system, when all subsystems have action of order $\hbar$, point structure of spacetime is lost.
There must exist a reformulation of quantum theory, which does not depend on classical spacetime.       [This becomes necessary if we remove all classical objects from today’s universe.]
This reformulation is a pre-quantum, pre-spacetime theory.
Such a pre-theory is in principle required at all energy scales, not just at the Planck energy scale, and carries more information than quantum field theory provides.

\section{Seven ingredients for unification}
The following seven ingredients are key to developing the sought for pre-spacetime, pre-quantum theory, and for proceeding from there to a theory of unification:

\begin{itemize}

\item  A new dynamics: trace dynamics [the work of Stephen L. Adler and collaborators] \cite{Adler}.

\item A new Lagrangian: based on the spectral action principle [the work of Chamseddine and Connes] \cite{CC}.

\item A new coordinate system: split bioctonions.

\item An old symmetry: $E_8 \times E_8$ and its branching.

\item Clifford algebras, spinors and particle content.

\item Exceptional Jordan algebra, and derivation of (some of the) dimensionless fundamental constants.

\item Recovering the observed universe: standard model + general relativity. And we predict two new forces, thus bringing the total number of fundamental forces to six.

\end{itemize}

In the following sections we describe these ingredients one by one,  very briefly.

\section{Part I. Adler's theory of trace dynamics}

In conventional quantisation in the Heisenberg picture, given a Hamiltonian dynamics, one raises classical dynamical variables to the status of matrices/operators.
Then we replace Poisson brackets by Heisenberg commutation relations,
and we replace Hamilton’s equations of motion by Heisenberg equations of motion.

Adler's trace dynamics is motivated by wanting to derive quantum (field) theory from a more fundamental theory, rather than by quantising classical dynamics.  Therefore, in trace dynamics, one does raise classical dynamical variables to matrices, but does not impose Heisenberg commutation relations on these matrices. Instead, one
constructs a   Lagrangian / Hamiltonian dynamics with matrix-valued dynamical variables. As an illustrative example, consider the action principle for a collection of non-relativistic free particles
\begin{equation}
S = \sum_i\;  \int d\tau \; \frac{1}{2}m_i\left(\frac{dq_i}{d\tau}\right)^2
\end{equation}
 Now, make a transition from real numbered dynamical variables $q$ to matrices $\bf q$ and propose the following new action:
 \begin{equation}
 S = \sum_i\;  \int d\tau \; Tr\left[ \frac{1}{2}\frac{L_{p}^2}{L^2}\left(\frac{d{\bf q}_i}{d\tau}\right)^2 \right]
\end{equation}
The key point is that the original Lagrangian turns into a matrix-valued polynomial, and by taking its matrix trace, a trace Lagrangian is constructed, and used to define the new action. A variational principle based on the `trace derivative' enables matrix-valued equations of motion to be derived, which in the present case are simply $\ddot{\bf q}_i =0$.

The Lagrangian, as well as the Hamiltonian, being a matrix trace, are invariant under global unitary transformations [provided no matrix-valued constant coefficients are present]. As a result, the Hamiltonian possesses a novel Noether charge, defined as follows
\begin{equation}\tilde{C} = \sum_i \; [{\bf q}_{Bi}, {\bf p}_{Bi}] -\left\{{\bf q}_{Fi},{\bf p}_{Fi}\right\}  \end{equation}
and known as the Adler-Millard charge. The subscript $B/F$ denotes a bosonic/fermionic degree of freedom. The dynamical matrices ${\bf q}$ have Grassmann numbers as their entries, and any $\bf q$ can be written as $\bf q = \bf q_B + \bf q_F$, where $\bf q_B / \bf q_F$ are Grassmann even / Grassmann odd. The presence of this charge, which has dimensions of action, makes trace dynamics into a pre-quantum theory: the sum of commutators over all degrees of freedom  is conserved, instead of each individual commutator being $i\hbar$.

\subsection {Trace dynamics $\longrightarrow$ quantum field theory as en emergent phenomenon}
Trace dynamics  is a Lorentz-invariant matrix-valued Lagrangian / Hamiltonian dynamics. It is deterministic, non-unitary but norm-preserving. Assume trace dynamics to hold at some time resolution $\tau_P$ at which quantum theory has not yet been tested in the laboratory. And ask, what is the emergent dynamics at a poorer time resolution $\tau \gg \tau_P$? There are two possibilities:
(i) In the approximation that  the Hamiltonian $\bf H$ is self-adjoint (and evolution is hence unitary), the conventional techniques of statistical thermodynamics are used to show that the novel charge is equipartitioned:
\begin{equation} [q_B, p_B]=i\hbar \ \ ; \quad \{q_F, p_F\} = i\hbar\end{equation}
 Heisenberg equations of motion and the laws of quantum field theory emerge at thermodynamic equilibrium.
 (ii) Second possibility: if the anti-self-adjoint part of the Hamiltonian is significant, evolution 
   breaks quantum linear superposition.
Coarse-graining the dynamics introduces an apparent randomness in the 
    evolution, while obeying Born rule : this is spontaneous localisation.
The underlying theory of trace dynamics is deterministic and non-unitary.
If non-unitarity is significant, linear superposition is broken.
We see that upon coarse-graining, the emergent theory is either quantum theory, or
    it is classical dynamics, depending on whether or not non-unitarity is significant.
    
    \subsection{Limitations of trace dynamics:} Trace dynamics does not specify any fundamental Lagrangian.
Also, it is a pre-quantum theory, not a pre-spacetime theory.
Spacetime is assumed classical, and flat (no gravitation). But this is an 
    intermediate step towards developing a pre-spacetime, pre-quantum theory.
We now propose to include matrix-valued gravitation, and a non-commutative
     pre-spacetime, and also propose a fundamental Lagrangian.
This leads to a candidate theory for unification of fundamental forces, which we refer to as
generalised trace dynamics. 

\section{Part II:  Generalised trace dynamics, and a Lagrangian}
We need to look for suitable classical dynamical variables for gravitation, which we can raise to the status of matrices / operators. The eigenvalues of the Dirac operator $D$ on a curved space-time manifold are related to the Einstein-Hilbert action as follows, in a regularised and truncated heat kernel expansion \cite{CC}:
\begin{equation}
Tr [L_P^2 D^2] \sim L_P^{-2}\int d^4x\; \sqrt{g}\; R + {\cal O}(L_P^{0}) =\sum_{i}\lambda_i^2 
\end{equation}
Therefore, the eigenvalues of the Dirac operator can play the role of dynamical variables in general relativity, instead of the metric \cite{Landi}. This turns out to have a deep connection with the fact that on a split biquaternionic / bioctonionic space, the Dirac operator is the same as the gradient operator on that space. It corresponds to the Klein-Gordon operator on 6D / 14D spacetime, with signature (3,3) and (7,7) respectively.

The sum over Dirac eigenvalues on the far right side of the above equation represents the Einstein-Hilbert action. For obvious reasons it is called the spectral action. We raise each eigenvalue $\lambda_i$ to the status of an operator / matrix $\hat{\lambda}_i\equiv \dot{q}_{Bi}$, thereby also introducing the matrix-valued bosonic gravitational variables $\dot{q}_{Bi}$. These are in fact dynamical variables for pre-quantum gravity; they give `curvature' to pre-spacetime. Each $\dot{q}_B$ is an `atom of space-time'. For each atom of space-time the trace dynamics action is obtained by mapping the contribution $\lambda_i$ in the classical action, to the trace action: $\lambda_i^2 \rightarrow \int d\tau\; Tr [\dot{q}_{Bi}^2]$. Here, $\dot{}$ denotes a time-derivative with respect to Connes time $\tau$, an absolute parameter resulting from the application of the 
Tomita-Takesaki theory to a non-commutative geometry.

Landi and Rovelli show how to write the Dirac action for a fermion in terms of the aforesaid Dirac eigenvalues. therefore giving it a spectral structure. Once again, each eigenvalue is raised to the status of a fermionic operator, denoted $\dot{q}_F$, and the following action for an `atom of space-time-matter' is proposed:
\begin{equation}
 \frac{S}{\hbar}= \int \frac{d\tau}{\tau_{Pl}}\; Tr \bigg\{ \frac{L_P^2}{L^2}\left[\dot{q}_B^\dagger +  \frac{L_P^2}{L^2}\beta_1 \dot{q}_F^\dagger \right ] \times \left[\dot{q}_B + \frac{L_P^2}{L^2} \beta_2 \dot{q}_F\right] \bigg\}
\end{equation}
Both sides of this equation are dimensionless, and Connes time has been scaled by Planck time. $\beta_1$ and $\beta_2$ are two unequal Grassmann numbers introduced to make the Lagrangian bosonic. The spectral action principle has been generalised so as to include Yang-Mills fields by the usual generalised Dirac operator $D\rightarrow D_A = D + A$
\begin{equation}
Tr [L_P^2 D_A^2] \sim L_P^{-2}\int d^4x\; \sqrt{g}\; R   + \frac{g_0^2}{N} \int_M d^4x \sqrt{g} g^{\mu\sigma}g^{\nu\rho}\partial_\mu A^i_\nu\partial_\sigma A^i_\rho  =\sum_{i}\lambda_{iA}^2
\end{equation}
By raising each eigenvalue $\lambda_{iA}$ to the status of an operator $\dot{q}_B + q_B$ we arrive at the final definition of an atom of space-time-matter [an STM atom or an `aikyon']
\begin{equation}
\frac{S}{\hbar}= \int \frac{d\tau}{\tau_{Pl}}\; Tr \bigg\{ \frac{L_P^2}{L^2}\left[\left(\dot{q}_B^\dagger + i\frac{\alpha}{L} q_B^\dagger\right) + \frac{L_P^2}{L^2}\beta_1\left( \dot{q}_F^\dagger + i\frac{\alpha}{L}q_F^\dagger\right)\right ] \times \left[\left(\dot{q}_B + i\frac{\alpha}{L}q_B^\dagger\right) + \frac{L_P^2}{L^2} \beta_2 \left(\dot{q}_F+ i\frac{\alpha}{L}q_F\right)\right] \bigg\}
 \label{aikyacn2}
\end{equation}
In our work, the $q_F$ describe left-chiral fermions, and the $\dot{q}_F$ describe right chiral fermions. Here, $\alpha$ is the dimensionless Yang-Mills coupling constant. An aikyon is an elementary particle such as an electron along with all the bosonic fields it produces. By summing over all STM atoms in our universe ($N\sim 10^{122})$ we arrive at the action principle for generalised trace dynamics. It contains all of the standard model as well as general relativity. Henceforth, we investigate the properties of just one aikyon, and the equations of motion can be obtained easily after first defining new variables
\begin{equation}
 q_1 = q_B + \beta_1 \frac{L_P^2}{L^2} q_F \ ;\qquad 
q_2 = q_B + \beta_2 \frac{L_P^2}{L^2} q_F  
\end{equation}
thus obtaining
\begin{equation}
\ddot q_1 = - \frac{\alpha^2}{L^2} q_1\ ; \qquad 
\ddot q_2 = - \frac{\alpha^2}{L^2} q_2
\end{equation}
The action can be simplified after first defining
\begin{equation}
{\dot{{Q}}_B} = \frac{1}{L} (i\alpha q_B + L \dot{q}_B); \qquad  {\dot{{Q}}_F} = \frac{1}{L} (i\alpha q_F + L \dot{q}_F);
\end{equation}
and
\begin{equation}
\dot{{Q}}_{1} ^\dagger   =   \dot{{Q}}_{B}^\dagger + \frac{L_{p}^{2}}{L^{2}} \beta_{1} \dot{{Q}}_{F}^\dagger  ; \ \qquad \dot {{Q}}_{2} =  \dot{{Q}}_{B} + \frac{L_{p}^{2}}{L^{2}} \beta_{2} \dot{{Q}}_{F}
\end{equation}
thus giving
\begin{equation}
\frac{S}{\hbar} = \frac{1}{2}\int \frac{d\tau}{\tau_{Pl}}\; Tr \biggl[\biggr. \frac{L_{p}^{2}}{L^{2}}  \dot{Q}_{1}^\dagger\;  \dot {Q}_{2} \biggr]
\end{equation}
This is the fundamental form of the action for an aikyon - the form it takes in the unification phase. The standard model aspect and the gravitation aspect are not yet manifest; in fact even the bosonic and fermionic aspects are not explicit in the unified phase. This is the phase before a left-right symmetry breaking (in our work this is the same as electro-weak symmetry breaking) which is precipitated by a quantum-to-classical transition resulting in the emergence of classical space-time. Each aikyon is described by a pair of dynamical variables ${Q}_1$ and $Q_2$, so that we may refer to an aikyon as a 2-brane (in the next section we suggest a pair of coordinates for the 2-brane). Each of these dynamical variables lives on a 16D split bioctonionic space, and the action will be assumed to have an $E_8 \times E_8$ symmetry.

\section{Part III. Quaternions and octonions as coordinates: a non-commutative pre-spacetime} 
In seeking to remove the real-number based point structure of spacetime from quantum field theory, we appeal to division algebras. There are only two more besides the reals and the complexes, these being the non-commutative quaternions and the non-commutative, non-associative octonions. The algebra automorphisms of the octonions play the role of unifying gauge transformations and spacetime diffeomorphisms. Octonionic space brings together the vector bundle (of gauge fields) and the spacetime manifold. 16D split bioctonionic space consists of two halves of eight each, and symmetry breaking segregates each of the eight components into a 4D quaternionic subspace which becomes spacetime, and the remaining 4D space is for the fibre bundle associated with an $SU(3)$ gauge symmetry. The gauge symmetry associated with the 4D spacetime is $SU(2) \times U(1)$. One 8D half of the split bioctonion (the left chiral half) associates with the standard model and a (second copy of) 4D spacetime curved by the weak force. The other half (the right chiral half) is the gravitational counterpart of the standard model. This latter also has an $SU(3) \times SU(2) \times U(1)$ gauge symmetry, with the $SU(3)$ being the gravitational analog of the strong force, and the remaining gauge sector is the precursor of general relativity and MOND.

The dynamical variable describing an aikyon will be assumed to live over a 16D space labeled by a split bioctonion, which we now introduce. Recall that a quaternion is a 4-d number system
\begin{equation}
q = a_0 + a_1 \hat i + a_2 \hat j + a_3 \hat k
\end{equation}
with a specified product rule amongst the imaginary directions, and an octonion is a 8-d number system
\begin{equation}
O = a_0 + a_1 {\bf e}_1 + a_2 {\bf e}_2 + a_3 {\bf e}_3 + a_4 {\bf e}_4 + a_5 {\bf e}_5 + a_6 {\bf e}_6 + a_7 {\bf e}_7
\end{equation}
with its own product rule. The 3D gradient operator associated with the imaginary space of the quaternion
\begin{equation}
\nabla= \hat{i}\frac{d\ }{dx} + \hat{j} \frac{d\ }{dy} + \hat{k} \frac{d\ }{dz}
\end{equation}
is a Dirac operator because its square is minus the Laplacian. An analogous result holds for the octonionic gradient operator in 7D. If we wish a gradient operator made from a division algebra to square to the Klein-Gordon operator, this is possible only in 6D spacetime (3,3) and in 14D spacetime (7,7), and results respectively from split biquaternions and split bioctonions.

Bosons and fermions live in octonionic space; more precisely, in split bioctonionic space, which enables the existence of chiral fermions. Just as in ordinary classical mechanics, position and momentum have vector components over 3D / 4D coordinates, here the matrix-valued dynamical variables have matrix components over octonionic space. For instance,
\begin{equation}
Q_B = Q_0 + Q_1 {\bf e}_1 + Q_2 {\bf e}_2 + Q_3 {\bf e}_3 + Q_4 {\bf e}_4 + Q_5 {\bf e}_5 + Q_6 {\bf e}_6 + Q_7 {\bf e}_7
\end{equation}

\subsection{Split biquaternions and split bioctonions}
A split complex number $\omega$ is defined as
\begin{equation}
\omega \ ; \quad \omega^2 = 1, \ ; \quad\omega\neq \pm 1, \ ; \quad w^* = -\omega, \; \quad z=a+\omega b
\end{equation}
and a split bicomplex number is $z + \omega \overline z$. The absolute magnitude of $z$ is $x^2 - y^2$, not $x^2 + y^2$, and is hence Lorentzian, not Euclidean. This is the reason why split biquaternions and split bioctonions are so important in our work. To define a split biquaternion recall that complex numbers generate the Clifford algebra $Cl(0,1)$ and the quaternions generate the Clifford algebra $Cl(0, 2)= (1, e_1, e_2, e_1e_2)$. The Clifford algebra $Cl(0,3)$
\begin{equation}
 Cl(0, 3)= (1, e_1, e_2, e_3, e_1e_2, e_2e_3, e_1e_3, e_1e_2e_3) \rightarrow (1, e_1, e_2, e_1e_2), \quad (e_1e_2e_3, e_2e_3, e_3e_1,e_3)
\end{equation} 
is generated by the split biquaternion \cite{Vaibhav}:
\begin{equation}
\rightarrow  (1, e_1, e_2, e_1e_2), \quad \omega (1, -e_1, -e_2, -e_1e_2)\qquad {\rm where}\ \omega=e_1e_2e_3 \rightarrow {\mathbf H} \oplus \omega {\mathbf H}\rightarrow {\mathbf C'}\otimes \mathbf H \rightarrow (1,\omega)\otimes \mathbf H
\end{equation}
Complex quaternions generate the algebra $Cl(2)$
\begin{equation}
Cl(2) = \mathbf C\otimes Cl(0,2) = \mathbf C \otimes \mathbf H
\end{equation}
Complex split biquaternions generate the algebra $Cl(3)$
\begin{equation}
Cl(3) = \mathbf C\otimes Cl(0,3) = \mathbf C \otimes (\mathbf H +\omega \mathbf H) \rightarrow (1,\omega)\otimes \mathbf H \sim Cl(2)\oplus Cl(2)
\end{equation}
and spinors constructed from them are adequate for describing chiral leptons.

Consider the entity characterised by the six imaginary directions of the complex split biquaternion (Clifford algebra $Cl(3)$; three vectors and three bivectors) as follows:
\begin{equation}
x_6 = it_1 \hat l + i t_2 \hat m + i t_3 \hat n + x_1 \hat i + x_2 \hat j + x_3 \hat k
\label{3ve6}
\end{equation}
where $i=\sqrt{-1}$. The magnitude of this vector is
\begin{equation}
|x_6|^2 = x_6 {\tilde x}_6 = {\tilde x}_6 x_6 = -t_1^2 - t_2^2 - t_3^2 + x_1^2 + x_2^2 + x_3^2
\label{3ve6sq}
\end{equation}
It describes a space-time interval in 6D spacetime with signature $(3,3)$ and having the symmetry group $SO(3,3)$. The Dirac operator $D_6$ is the gradient operator
\begin{equation}
D_6 = i\hat l\frac{ \partial\ }{\partial t_1} + i\hat m \frac{ \partial\ }{\partial t_2} + i\hat n\frac{ \partial\ }{\partial t_3} + \hat i \frac{ \partial\ }{\partial x_1} + \hat j\frac{ \partial\ }{\partial x_2} + \hat k\frac{ \partial\ }{\partial x_3} 
\label{3D6op}
\end{equation}
and the square of $D_6$ gives the Klein-Gordon operator
\begin{equation}
D_6^2 = -\frac{\partial\ }{\partial t_1^2}  -\frac{\partial\ }{\partial t_2^2}  -\frac{\partial\ }{\partial t_3^2}  +\frac{\partial\ }{\partial x_1^2}  +\frac{\partial\ }{\partial x_2^2}  +\frac{\partial\ }{\partial x_3^2} 
\label{3D6sq}
\end{equation}
This 6D spacetime splits into two copies of 4D spacetime with relatively flipped signatures \cite{Kritov}. Geometry of our spacetime is given by gravitation via the general theory of relativity [$SU(2)_R \times U(1)_{YDEM}$]. Whereas the geometry of the second 4D spacetime is given by the weak force [$SU(2)_L \times U(1)_Y$]. The weak force is left-handed gravity: it is a spacetime symmetry masquerading as an internal symmetry! This helps understand why weak force violates parity; so should gravitation. The two 4D spacetimes share one time and one space direction. In other words, the two additional dimensions beyond 4D are timelike, not spatial. Experiments needed to probe the second spacetime will have to be at the weak interaction length scale $10^{-17}$ cm \cite{Kau}.

Complex octonions generate the Clifford algebra $Cl(6)$
\begin{equation}
Cl(6) = \mathbf C\otimes Cl(0,6) = \mathbf C \otimes \mathbf O
\end{equation}
Complex split bioctonions generate the Clifford algebra $Cl(7)$
\begin{equation}
Cl(7) = \mathbf C\otimes Cl(0,7) = \mathbf C \otimes (\mathbf O +\omega \mathbf O) \sim Cl(6)\oplus Cl(6)\; , \qquad \omega = \overleftarrow{e_1e_2e_3e_4e_5e_6e_7}
\end{equation}
and these are required if one wants to describe chiral quarks as well as chiral leptons. The 14D gradient operator made from the fourteen imaginary directions of the split bioctonion describes 14D spacetime with signature (7,7). For deep reasons which remain to be fully understood, left-right symmetry breaking sends this to the 6D spacetime generated by the complex split biquaternions whereas the remaining eight directions (Euclidean signature) serve to define two copies of 4D vector bundles associated with the unbroken gauge symmetry $SU(3)_{color}\times SU(3)_{grav}$. We tentatively suggest that the two real directions associated with the split bioctonion serve to define the 2D space for the 2-brane defining the aikyon's dynamical variable.

The five exceptional Lie groups $G_2, F_4, E_6, E_7$ and $E_8$, along with the Freudenthal-Tits magic sqaure, play an important role in our research programme. In the magic square, the last four of these groups are respectively expressed as $3\times 3$ matrices with entries in tensor products over the division algebras $\mathbf R\otimes \mathbf O, \mathbf C\otimes \mathbf O, \mathbf H\otimes \mathbf O, \mathbf O\otimes \mathbf O$. Also significant is the relation between division algebras and Lorentz groups: 
\begin{equation}
SL(2,\mathbf R) \sim SO(1,2); \qquad SL(2,\mathbf C) \sim SO(1,3)\ ; \quad SL(2,\mathbf H) \sim SO(1,5)\ ; \quad SL(2,\mathbf O) \sim SO(1,9)
\end{equation}

With the introduction of the non-commutative pre-spacetime for every aikyon, we now have the sought for pre-quantum, pre-spacetime theory, in the form of generalised trace dynamics. Remarkably, with the choice of an appropriate gauge group, it is also a theory of unification. The gauge group is in fact dictated by the octonions themselves, and moreover, spinors made from Clifford algebras derived from quaternions and octonions describe fermionic states. We see that once we appeal to division algebras, they generate enough mathematical structure (algebraic as well as geometrical) to cover various aspects of unification.

\section{Part IV.  $E_8\times E_8$ unification of standard model and pre-gravitation }

  The origins of the $E_8 \times E_8$ theory lie in quantum foundations. There ought to exist a reformulation of quantum field theory in which space-time is not described by real numbers \cite{Singh1}. This requirement is a must at all energy scales, including those currently accessed by particle colliders, and not just at the Planck scale. The current formulation of quantum field theory (QFT) based on a classical spacetime manifold, despite its enormous success, must be treated as approximate. This can be appreciated by noting that the theory does not explain why the standard model has these specific gauge symmetries, and why the dimensionless coupling constants of the standard model take the seemingly arbitrarily values measured in experiments. Some researchers would say that answers to these questions lie in a yet to be discovered unified theory arising in the vicinity of the Planck scale. However, we have argued \cite{Singh1} that the answers to these questions can be found at low energies itself, by generalising QFT to a formulation from which the classical spacetime with its commutative point structure is removed. Instead, we label pre-spacetime coordinates by the non-commutative and non-associative division algebraic generalisations of reals and of complex numbers, such generalisations being the quaternions and the octonions. These generalisations unify spacetime diffeomorphisms and gauge transformations, with the quaternions being adequate for unifying general relativity with the electroweak interaction. (The split complex number $\omega$ is what enables a non-compact Lorentzian structure  to emerge from a tensor product such as $SU(2) \times \omega SU(2)$). The octonions are essential for bringing in the strong force. More precisely, the split bioctonions are required, so as to have chiral fermions. The use of $E_8$, the largest of the five exceptional Lie groups, in the tensor product form $E_8 \times E_8$, is found to admit the observed chiral fermions, the standard model gauge symmetries, general relativity, the dark electromagnetic field $U(1)_{grav}$ proposed as the origin of MOND, and a sixth new force, $SU(3)_{grav}$, this being the gravitational analog of the strong force. Two of the $SU(3)$ groups in the branching are responsible for the emergence of 6D spacetime \cite{Kau} as well as the internal symmetry  gauge space. Equally importantly, the theory provides strong evidence that several of the dimensional fundamental constants can be derived from first principles \cite{Singh1}.

Why this particular branching, in which each of the two $E_8$-s branches into four $SU(3)$s? This is motivated from the celebrated Freudenthal-Tits magic square in which the exceptional Lie groups (except $G_2$) are expressed as tensor products of matrices over two division algebras. One of the two algebras is that of the octonions, and the second one is respectively reals, complexes, quaternions and octonions, for giving rise to the groups $F_4, E_6, E_7, E_8$. Starting with $E_8$ we ask what is the decomposition which preserves the complex structure? The answer is $SU(3) \times E_6$. Next, what is the decomposition of $E_6$ which preserves its complex structure? The answer is $SU(3) \times F_4$. The complex structure of $F_4$ is preserved by the decomposition $SU(3) \times SU(3)$. This is how the proposed branching is arrived at. The six  emergent interactions, resulting from a left-right symmetry breaking (same as the electroweak symmetry breaking) are
\begin{equation}
SU(3)_C \otimes \; SU(2)_{L} \otimes \; U(1)_Y \otimes \; SU(3)_{grav} \otimes\; SU(2)_{R} \otimes\; U(1)_{g}
\end{equation}

The ${\bf 248}$ fundamental rep of $E_8$ is remarkably also its adjoint rep. The $(\bf 248, \bf 1) \oplus (\bf 1, \bf 248)$ is broken into two separate $E_8$. The first  $E_8$ branches into  $SU(3)_{space1} \otimes E_6$, and analogously the second $E_8$ branches into $SU(3)_{space2} \otimes E_6$. Each of the $E_6$ undergoes the well-known trinification
\begin{equation}
E_6\rightarrow SU(3) \otimes SU(3) \otimes SU(3)\; , \quad \bf 248 = (8,1) \oplus (1,78) \oplus (3,27) \oplus (\bar{3}, \overline{27})
\end{equation}
The first $E_6$ gives the standard model: $E_6 \rightarrow SU(3)_{LHgen} \otimes SU(3)_c \times SU(2)_L \times U(1)_{Y}$ whereas the second $E_6$ gives its pre-gravitational counterpart: $ E_6 \rightarrow SU(3)_{RHgen} \otimes SU(3)_{grav} \times SU(2)_R \times U(1)_{g}$.

\subsection{Particle content after $E_8 \times E_8$ branching:}
The particle content arising after symmetry breaking is shown in Figs. 1 and 2. The quarks are leptons of the standard model are reproduced with correct charges, along with three types of right-handed sterile neutrinos \cite{Kau}. We have a left-right symmetric theory originally, and hence the electroweak breaking is a chiral symmetry breaking.

\begin{figure}[t]
\begin{center}
\includegraphics[scale=.25]{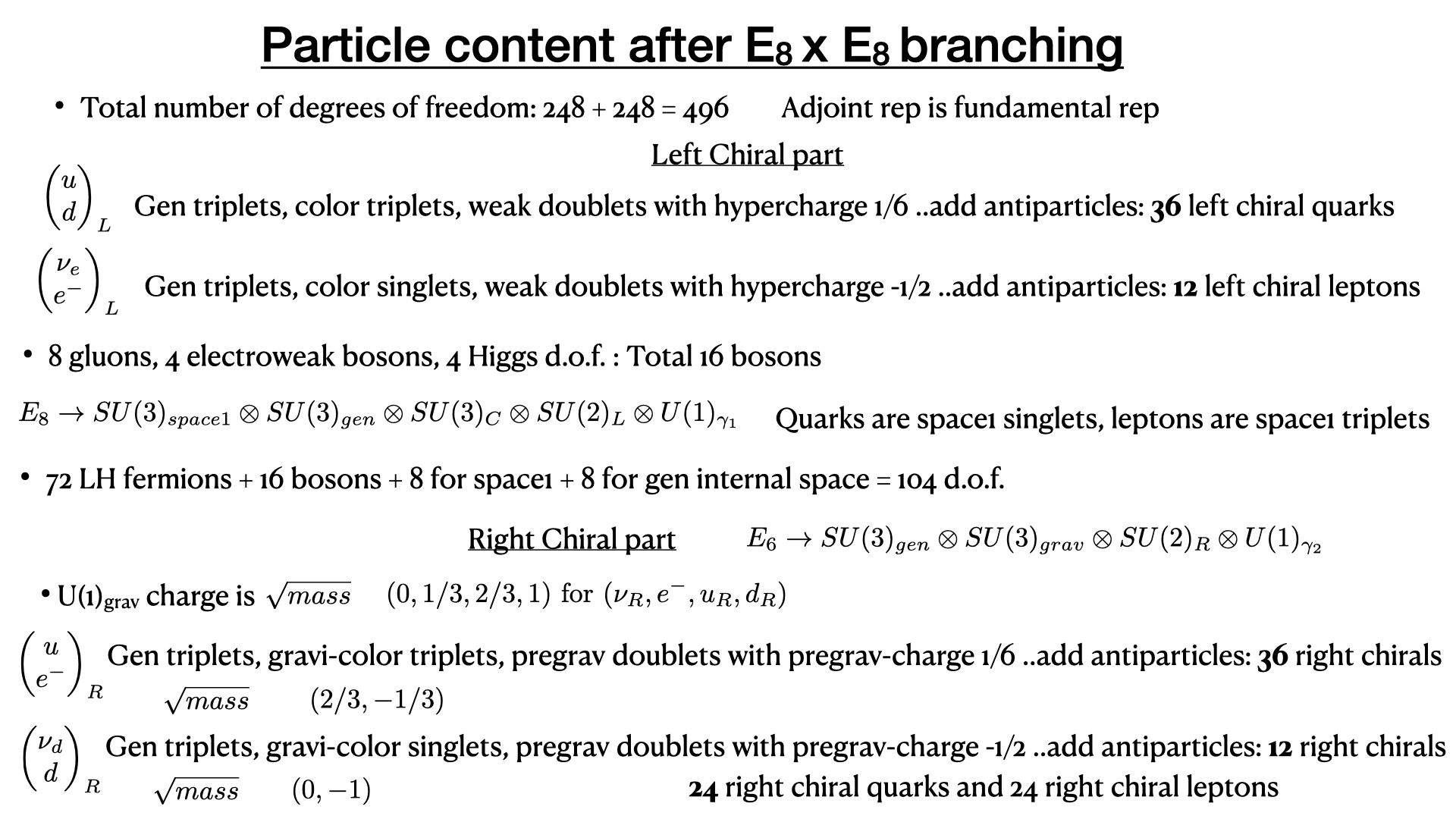}
%
%
\caption{The particle content of the theory - part I.}
\label{4fig1}       
\end{center}
\end{figure}

\begin{figure}[t]
\begin{center}
\includegraphics[scale=.25]{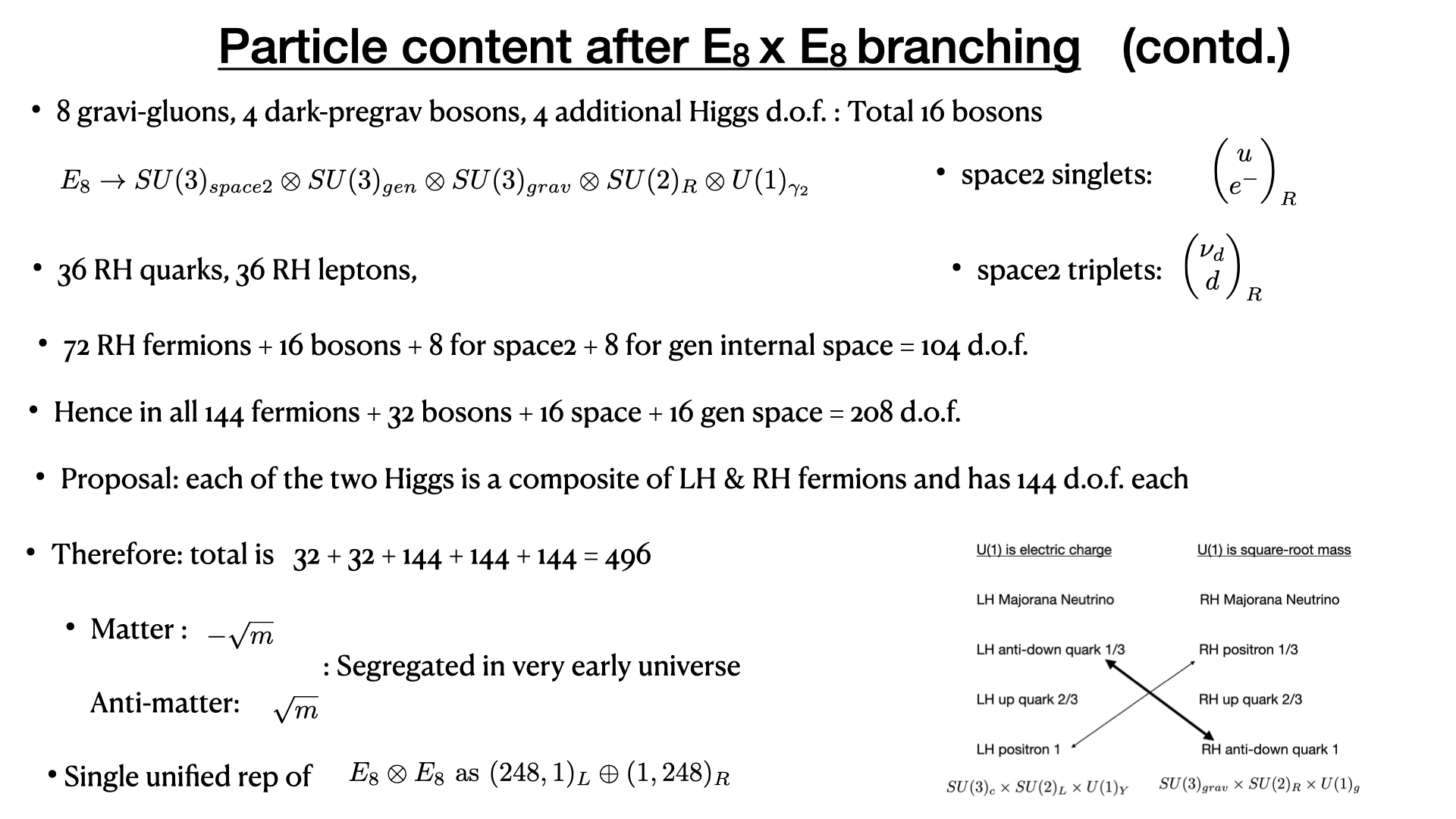}
%
%
\caption{The particle content of the theory - part II.}
\label{4fig2}       
\end{center}
\end{figure}

One key new feature of our theory is that we have introduced the charge square-root of mass for $U(1)_{grav}$, in analogy with electric charge for $U(1)_{em}$. The motivation for introducing the charge $\pm \sqrt m$ is the experimental fact that the square roots of the masses of the electron, up quark, down quark are in the ratio $(1/3, 2/3, 1)$. This should be contrasted with their electric charge ratios $(1, 2/3, 1/3)$ which remarkably happen to be the same across all three generations whereas the mass ratios for the second and third generation show no obvious pattern. In order to be able to explain this puzzle, we proposed $\pm \sqrt m$ as the charge for $U(1)_{grav}$ (for all three generations) and interchanged the relative position of the right handed electron and right handed down quark (as also those of their heavier counterparts)  in their color assignment for $SU(3)_{\rm grav}$. The electron is a triplet of $SU(3)_{\rm grav}$ and the down quark is a singlet. This enabled us to theoretically derive the strange mass ratios for the second and third generation \cite{Bhatt},  and in an analogous derivation, also the parameters in the CKM quark mixing matrix \cite{ckm}.

 $U(1)_{grav}$, which has the desired properties of the dark matter fluid, is sourced by square-root of mass $\pm\sqrt{m}$. The plus sign is for matter and minus sign for anti-matter. Like signs attract and opposite signs repel, under the dark electromagnetic force. Our universe, being matter dominated as opposed to anti-matter dominated, largely has only $+\sqrt m$ particles. Thus for all practical purposes, dark electromagnetism in our universe is an attractive only force, unlike ordinary electromagnetism.
 
 Earlier, we have already mentioned the fundamental Lagrangian, in Section IV. The bosonic part of this Lagrangian, corresponding to the above branching, has been analysed by us recently in some detail. Interestingly, this analysis also yields a satisfactory derivation of the weak mixing angle \cite{Raj}.

\section{Part V. Clifford algebras and particle content}
As we have seen, the complex quaternions generate the Clifford algebra $Cl(2)$ and complex split biquaternions generate the Clifford algebra $Cl(3)$. Octonionic chains generate the Clifford algebra $Cl(6)$ when one out of the seven imaginary directions is kept fixed. The associated automorphisms form the subgroup $SU(3)$ of the automorphism group $G_2$ of the octonions. The action of $Cl(6)$ on an idempotent (constructed from acting the octonion algebra onto itself) gives rise to eight spinor states which possess the correct charges to describe one generation of standard model quarks and leptons under the unbroken symmetry $SU(3)\times U(1)$. More precisely, the left-chiral fermionic states are obtained by acting on the left handed neutrino and the associated unbroken symmetry is $SU(3)_{color} \times U(1)_{em}$. Right-chiral fermionic states are constructed by acting on the right-handed sterile neutrino and the associated gauge symmetry is $SU(3)_{grav}\times U(1)_{grav}$. Here, the right-handed down quark as well as the right-handed sterile neutrino are singlets of $SU(3)_{grav}$, whereas the right-handed up quark and electron are triplets. 

Figure 4 explicitly shows an algebraic construction of one generation of standard model quarks and leptons. Now, in principle, this can be any one of the three experimentally observed fermion generations, not necessarily the lightest one. Because apart from masses, these three generations are identical in all other properties (lepton universality) in so far as the SM is concerned. And yet, this algebraic construction picks out the lightest generation, because only for this generation the observed square-root mass ratios $(0,1/3,2/3,1)$ coincide with the mass ratios predicted by this construction. How then are we to explain the strange observed mass ratios for the other two generations, without contradicting the lepton universality expected from this algebraic construction? We answer this question in the next section.

The role of Clifford algebras as also of sedenions, and the deep question of why three fermion generations, is addressed in this proceedings volume in the article by Gresnigt and Gourlay \cite{Gresnigt}.

\begin{figure}[t]
\begin{center}
\includegraphics[scale=.25]{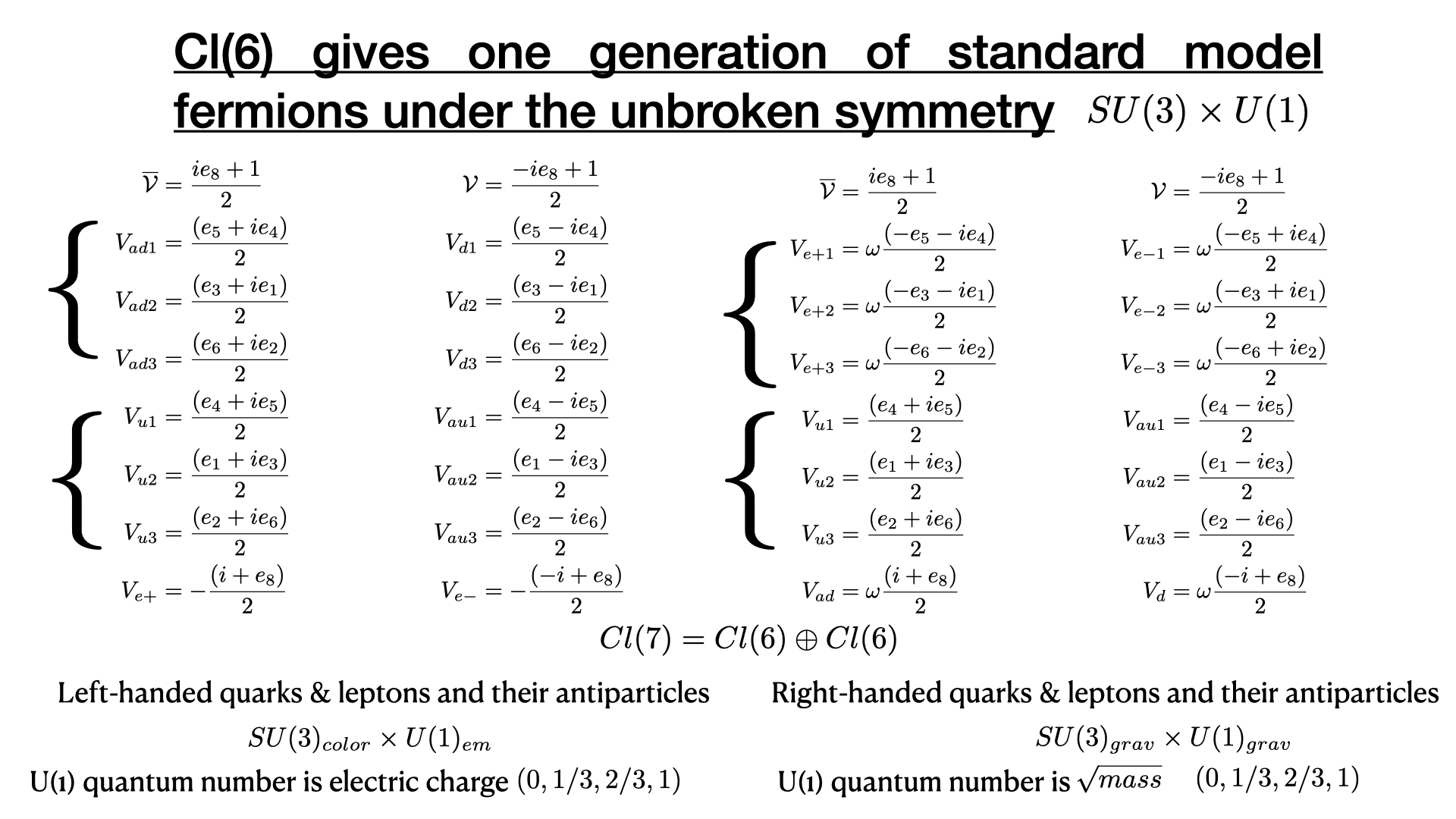}
%
%
\caption{Clifford algebras and fermionic particle states}
\label{4fig3}       
\end{center}
\end{figure}

\section{Part VI. Exceptional Jordan algebra and
derivation of fundamental constants}
A Jordan algebra  is defined as follows: given two elements $J_1$ and $J_2$ of a Jordan algebra, their Jordan product is defined by $J_1\circ J_2 = (J_1J_2 + J_2 J_1)/2$. Elements of Jordan's algebra also satisfy $J_1^2\circ (J_1\circ J_2)  = J_1\circ (J_1^2\circ J_2)$ known as the Jordan identity. The Jordan product is obviously commutative. It is not associative, but is power associative. Moreover, unlike for ordinary matrix multiplication of Hermitian matrices, the Jordan product of two Hermitian matrices is another Hermitian matrix.

In view of the definition of a Jordan algebra, a classical Lie group can be regarded as one that obeys the Jordan product, and these are hence automorphism groups of the corresponding algebra. Clearly, for the reals, complexes and quaternions, these are $SO(n), SU(n)$ and $Sp(2n)$. In going from the second column to the first column of the magic square the symmetric matrices that  are present are $SO(3)$, $SU(3)$ and $Sp(6)$, these being automorphism groups of the corresponding Jordan algebras. The same is true of the $3\times 3$ Hermitian matrix $F_4$, having octonionic entries. The automorphism group of such matrices is indeed $F_4$, as was shown by Chevalley and Schafer in 1950 \cite{Chevalley}. The algebra is known as the exceptional Jordan algebra (also known as Albert algebra); exceptional because its automorphism group is exceptional, and because  for octonions only $3\times 3$ Hermitian matrices form a Jordan algebra, no other.

Why is the exceptional Jordan algebra of physical interest? In 1934 Jordan,  von Neumann and Wigner published their famous paper  ``On an Algebraic Generalisation of the Quantum Mechanical Formalism" \cite{jvw}. 
Building on earlier work by Jordan, they investigated an algebra of observables defined by the Jordan product.
The property of power associativity ensures that if some observable $A$ takes the value $a$ then a function $f(a)$ of $a$ takes the value $f(a)$. These physically desirable features make the Jordan algebraic formulation of quantum mechanics an attractive alternative to the conventional formulation.
The Jordan algebra can also be directly defined as a vector space with a commutative bilinear product obeying the Jordan identity. And after defining a multiplicative identity element, it serves as an algebra of observables.

In their paper, Jordan et al. showed that this algebra is equivalent to the ordinary matrix algebra of Hermitian matrices (and hence equivalent to conventional quantum mechanics) of arbitrary dimension, when the matrix entries are real numbers, complex numbers, or quaternions. However, they found one exception to this equivalence, that being the Jordan algebra of $3\times 3$ Hermitian matrices with octonions as entries. This is the exceptional Jordan algebra (EJA) (also called the Albert algebra) and denoted $J_3(8)$. The authors conclude that `essentially new results, in contrast to the present content of quantum mechanics' are only to be expected in this one exceptional case. Several decades later, Townsend \cite{Townsend}
aired the same sentiment: ``The octonionic quantum mechanics based on the exceptional Jordan algebra 
 is the most interesting case because
exceptionality implies that no Hilbert space formulation is possible. This is therefore
a genuine, and radical, generalisation of quantum mechanics, although one without as
yet any application." What we have found in our unification programme is that there indeed is an application of significance, and that the characteristic equation of the EJA determines  values of the dimensional fundamental constants. This will be discussed below.

A typical matrix in the EJA has the form
\begin{equation}
J( x, \xi) = \begin{pmatrix} \xi_1 \ & x_3 \ & x_2^\dagger \\ x_3^\dagger\ & \xi_2 \ & x_1 \\ x_2 \ & x_1^\dagger \ & \xi_3\end{pmatrix}
\label{3ejamat}
\end{equation}
where $x_1, x_2, x_3$ are octonions, $\xi_1, \xi_2, \xi_3$ are reals, and dagger denotes octonionic conjugation. The automorphism group preserves the quadratic form ${\rm Tr} (J_1\circ J_2)$ (which is non-degenerate: ${\rm Tr} (J_1\circ J)=0\  \forall J\implies J_1=0$) and it also preserves the cubic form ${\rm Tr} [J_1\circ (J_2\circ J_3)]$. This makes the EJA crucially different from classical Lie algebras, which are defined only by the invariance of quadratic forms. The physical significance of the cubic form still remains to be unraveled, in our unification programme. 

The Freudenthal product of two Jordan matrices $J_1$ and $J_2$ is defined as
\begin{equation}
{J_1*J_2} = J_1\circ J_2 - \frac{1}{2} \left[ J_1{\rm Tr} (J_2) + J_2{\rm Tr} (J_1)\right] + \frac{1}{2}\left[{\rm Tr}(J_1){\rm Tr}(J_2) - {\rm Tr}(J_1\circ J_2) \right]
\end{equation}
The determinant of a Jordan matrix is defined as
\begin{equation}
{\rm det} (J) =\frac{1}{3}{\rm Tr} \left((J*J)\circ J\right)
\end{equation}
which, in terms of the components of $J$ shown in (\ref{3ejamat}) becomes
\begin{equation}
{\rm det} J = \xi_1\xi_2\xi_3 + [x_3(x_2 x_1)] + [x_3(x_2x_1)]^\dagger - \xi_1 |x_1|^2 - \xi_2 |x_2|^2 - \xi_3 |x_3|^2
\end{equation}
The Jordan product and the Freudenthal product can be respectively thought of as generalisations of the ordinary dot product and cross product of vectors in $\mathbf R^3$. The exceptional Jordan algebra can be regarded as a generalisation of the vector space $\mathbf R^3$ to the case of octonions. The group $F_4$ is the group of transformations $J\mapsto MJM^\dagger$ which preserve the determinant and trace of a matrix in the EJA and can be justifiably named $SU(3,\mathbf O)$ as seen also in the magic square. For an elegant construction of $F_4$ using triality and three copies of $SU(2, \mathbf O)$ see Dray and Manogue, Section 11.3 \cite{Dray}.

From the magic square one can infer that the exceptional groups are a natural generalisation of orthogonal (including Lorentzian case), unitary and symplectic groups to the case of division and split algebras involving octonions. Clearly then, the exceptional groups represent a new geometry, which will be the geometry of the unified $E_8 \times E_8$ theory. The new geometry naturally unifies space-time geometry with the internal geometry of gauge interactions.

The characteristic equation of the EJA is
\begin{equation}
\lambda^3-Tr(X)\lambda^2+S(X)\lambda-Det(X)=0
\end{equation}
where
\begin{equation}
Tr(x)=\xi_1+\xi_2+\xi_3\ ,\ \ \ \  Det(X)=\xi_1 \xi_2 \xi_3 +2\ Re(x_1 x_2 x_3)-\sum\limits_{i=1}^3 \xi_i x_i \tilde{x}_i
\end{equation}
and
\begin{equation}
S(x)=\xi_1 \xi_2+\xi_2 \xi_3+\xi_3 \xi_1-x_1\tilde{x}_1-\tilde{x}_2x_2-x_3\tilde{x}_3
\end{equation}
In our work we have reasoned that the EJA is the correct place for describing three generations of a given fermion, with the diagonal entries (all three are equal) representing either electric charge or square-root mass. For left-handed fermions it is electric charge, and for right-handed fermions it is square root of mass. For the left-handed case the corresponding four matrices are (upto automorphisms, these four are the only ones possible).

\begin{equation}
 X_\nu = \begin{bmatrix}
                0 & V_{\tau} & \tilde{V_{\mu}}\\
                \tilde{V_{\tau}} & 0 & V_{\nu}\\ 
                V_{\mu} & \tilde{V_{\nu}} & 0
            \end{bmatrix}\quad
        X_{e^+} =  \begin{bmatrix}
                {1}{} & V_{a\tau} & \tilde{V_{a\mu}} \\
                \tilde{V_{a\tau}} & {1}{} & V_{e^+} \\ 
                V_{a\mu} & \tilde{V_{e^+}} & {1}{}
            \end{bmatrix}\quad
    X_{u} =\begin{bmatrix}
                \frac{2}{3} & V_{t} & \tilde{V_{c}} \\
                \tilde{V_{t}} & \frac{2}{3} & V_{u} \\ 
                V_{c} & \tilde{V_{u}} & \frac{2}{3}
            \end{bmatrix}\quad
    X_{ad} = \begin{bmatrix}
               \frac{1}{3}  & V_{ab} & \tilde{V_{as}} \\
                \tilde{V_{ab}} & \frac{1}{3} & V_{ad} \\ 
                V_{as} & \tilde{V_{ad}} & \frac{1}{3}\\
            \end{bmatrix}
\end{equation}
\begin{figure}[t]
\begin{center}
\includegraphics[scale=.25]{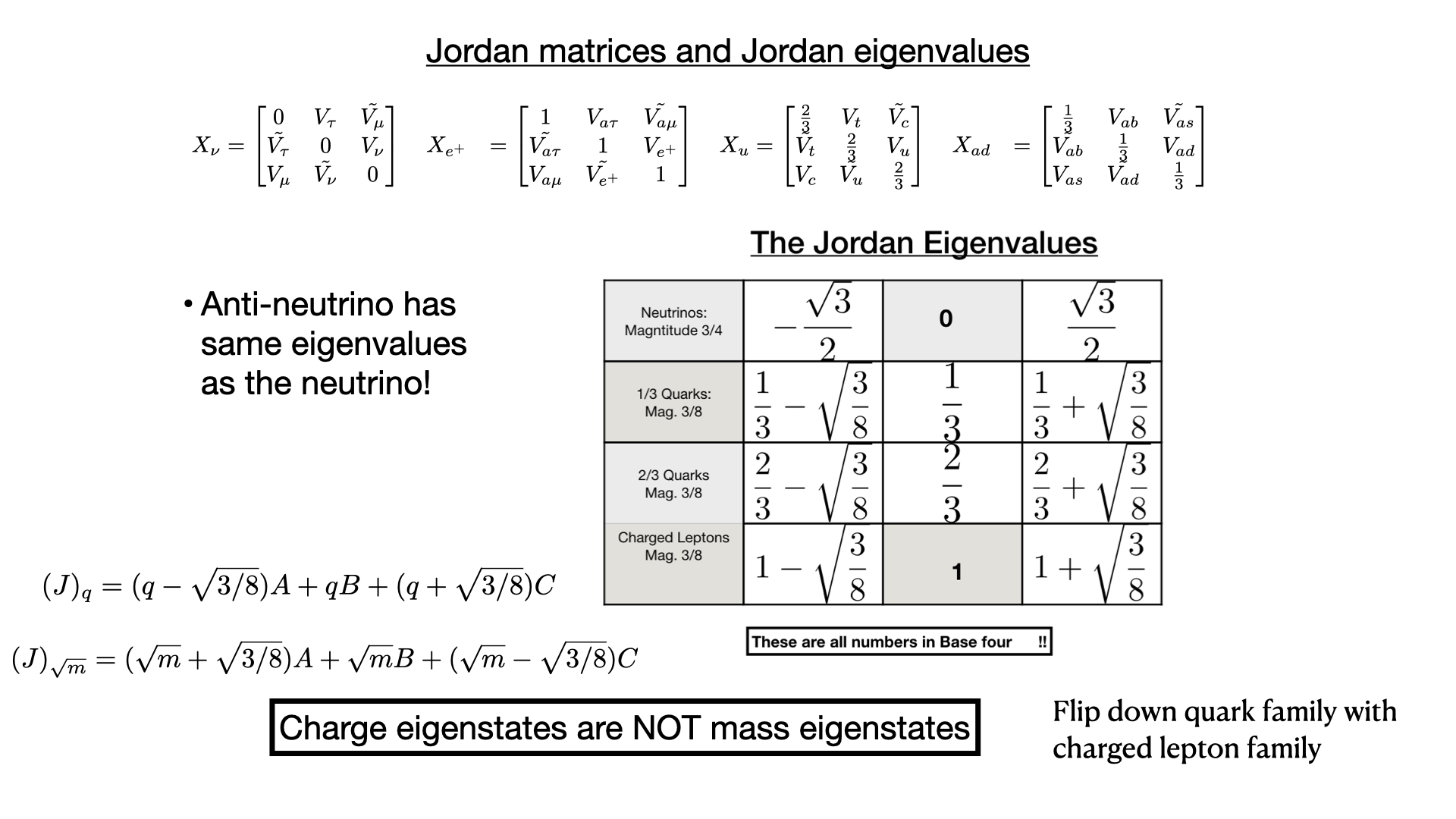}
%
%
\caption{The Jordan eigenvalues}
\label{4fig3}       
\end{center}
\end{figure}
At the heart of our understanding of mass ratios is the realisation that electric charge eigenstates (left chiral) are distinct from square-root mass eigenstates (right chiral). And that for all three fermion generations of right-chiral fermions, the square-root mass ratios are the same: $(0,1/3, 2/3,1)$, just as for electric charge. However, since in experiments all mass measurements are eventually electromagnetic measurements (which use charge eigenstates), square-root mass eigenstates must be expressed as superposition of charge eigenstates. The square-root mass ratios deduced therefrom are no longer $(0,1/3, 2/3,1)$ but are in fact the observed `strange' ratios \cite{Bhatt}.

The solution of the characteristic equation of the EJA permits us to write the electric charge eigenstates as superpositions of basis states $(A, B, C)$ (of  electric-charge-square-root-mass $e\sqrt{m})$ that exist prior to the left-right symmetry breaking (see Fig. 4). Thus,  the charge eigenstates $(J)_q$ are left-handed projections of the basis states, and the Jordan eigenvalues are coefficients in this projection. Similarly, the square-root mass eigenstates $(J)_{\sqrt m}$ are right-handed projections of the same basis states $(A, B, C)$, as shown in Fig. 4. Using these results, we can write square-root mass eigenstates as superpositions of charge eigenstates, and therefrom arrive at the experimentally observed mass ratios \cite{Bhatt}.

\section{Part VII. Emergence of classical spacetime and gravitation}
\begin{figure}[H]
\begin{center}
\includegraphics[scale=.20]{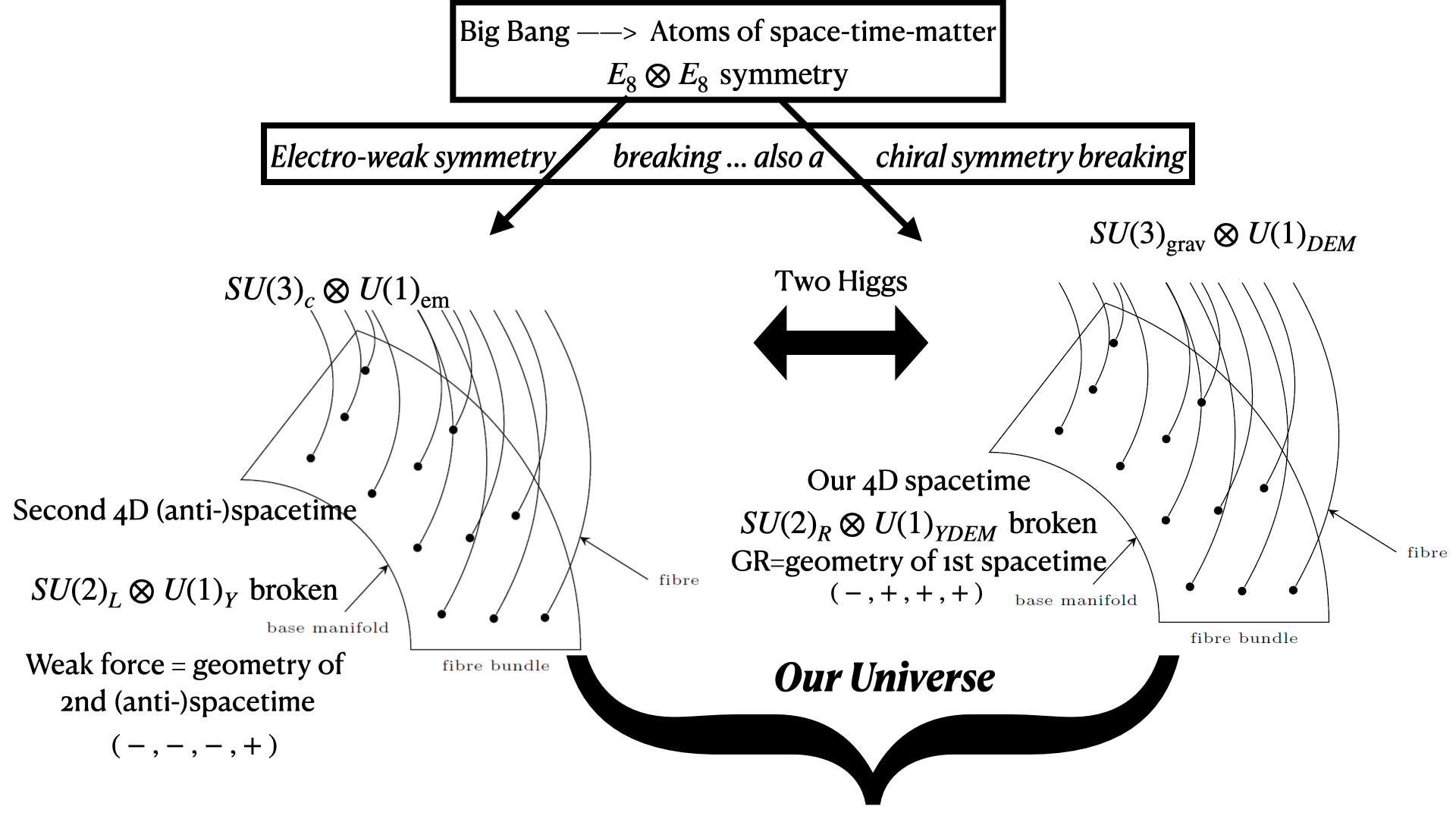}
%
%
\caption{The proposed emergence of our universe from the breaking of $E_8 \times E_8$ symmetry.}

\label{4figem}       
\end{center}
\end{figure}

Fig. \ref{4figem} shows the low energy emergent universe arising after electroweak symmetry breaking. Our universe consists also of  a second copy of 4D spacetime, with a signature flipped relative to ours: it has three timelike and one spacelike dimension. One time and one space direction is shared with ours, so that together they constitute a 6D spacetime with signature (3,3). From the vantage point of our 4D spacetime, the two extra time dimensions can be thought of as being in the vector bundle, and represent the geometry of the weak force. Only quantum systems access the 6D spacetime; classical systems live in 4D. The second 4D spacetime is curved by the weak force.

\end{document}